\documentclass[letterpaper]{article} 
\usepackage{aaai25}  
\usepackage{times}  
\usepackage{helvet}  
\usepackage{courier}  
\usepackage[hyphens]{url}  
\usepackage{graphicx} 
\usepackage{natbib}  
\usepackage{caption} 
\usepackage{algorithm}
\usepackage{algorithmic}

\usepackage{booktabs}
\usepackage{multirow}
\usepackage{amssymb}
\usepackage{amsmath}
\usepackage{newfloat}
\usepackage{listings}
\DeclareCaptionStyle{ruled}{labelfont=normalfont,labelsep=colon,strut=off} 
\floatstyle{ruled}
\newfloat{listing}{tb}{lst}{}
\floatname{listing}{Listing}
\title{Generative Medical Segmentation}
\author{
    Jiayu Huo\textsuperscript{\rm 1}\thanks{Corresponding author}
    Xi Ouyang\textsuperscript{\rm 2},
    S\'{e}bastien Ourselin\textsuperscript{\rm 1},
    Rachel Sparks\textsuperscript{\rm 1}
}
\affiliations{
    \textsuperscript{\rm 1}School of Biomedical Engineering and Imaging Sciences (BMEIS), \\
King's College London, London, UK\\

    \textsuperscript{\rm 2}Shanghai United Imaging Intelligence Co., Ltd., Shanghai, China\\
    jiayu.huo@kcl.ac.uk
}

\usepackage{bibentry}

\begin{document}

\maketitle

\begin{abstract}
Rapid advancements in medical image segmentation performance have been significantly driven by the development of Convolutional Neural Networks (CNNs) and Vision Transformers (ViTs). These models follow the discriminative pixel-wise classification learning paradigm and often have limited ability to generalize across diverse medical imaging datasets. In this manuscript, we introduce Generative Medical Segmentation (GMS), a novel approach leveraging a generative model to perform image segmentation. Concretely, GMS employs a robust pre-trained vision foundation model to extract latent representations for images and corresponding ground truth masks, followed by a model that learns a mapping function from the image to the mask in the latent space. Once trained, the model generates an estimated segmentation mask using the pre-trained vision foundation model to decode the predicted latent representation back into the image space. The design of GMS leads to fewer trainable parameters in the model which reduces the risk of overfitting and enhances its generalization capability. Our experimental analysis across five public datasets in different medical imaging domains demonstrates GMS outperforms existing discriminative and generative segmentation models. Furthermore, GMS is able to generalize well across datasets from different centers within the same imaging modality. Our experiments suggest GMS offers a scalable and effective solution for medical image segmentation. GMS implementation and trained model weights are available at \url{https://github.com/King-HAW/GMS}
\end{abstract}

%

\section{Introduction}
Image segmentation plays a critical role in medical image analysis by enabling the automated and precise delineation of anatomical and pathological structures within medical images. This process allows clinicians to obtain detailed visualizations, such as lesions and tumors, which can support computer-aided diagnosis systems and enhance the accuracy of clinical assessments. Additionally, the quantitative assessments derived from segmentation are vital for treatment planning and monitoring of disease progression. By incorporating automated segmentation into clinical practice, the precision and efficacy of therapeutic interventions are improved, leading to better patient outcomes~\cite{al2020dataset,tschandl2018ham10000,jha2021kvasir}.


Current deep learning models designed for medical image segmentation, such as U-Net~\cite{ronneberger2015u} and its various adaptations~\cite{ruan2023ege,ibtehaz2023acc}, have significantly advanced the field of medical imaging analysis. These models have been pivotal in enhancing the accuracy and efficiency of segmenting objects from various imaging modalities such as MRI and CT. Early deep learning-based image segmentation models leverage convolution kernels to learn local patch representations from large amounts of labeled data. Despite their successes, models based on Convolutional Neural Networks (CNNs) often have a large number of parameters which can introduce challenges in model training and increase the likelihood of overfitting when training datasets are small. Additionally, the limited receptive field of the convolution kernel makes it difficult for CNN-based models to learn global context information that can provide important guidance during image segmentation. Moreover, CNN-based models struggle with generalizing to unseen domains, leading to potentially substantial performance drops when the test dataset distribution is shifted from the training dataset distribution.

The Vision Transformer (ViT)~\cite{dosovitskiy2021an} model has recently been presented as a powerful alternative to CNN-based segmentation models in medical imaging analysis. ViT can capture global semantic information that the convolution kernel is unable to represent. Transformer-based segmentation models, such as UCTransNet~\cite{wang2022uctransnet} and Swin-Unet~\cite{cao2022swin}, leverage the transformer architecture to represent images as sequences of patches, enabling the model to learn relationships across the entire image. Transformer-based models facilitate a more holistic image analysis by integrating both local and global context information. Therefore, these models can accurately segment anatomical structures or pathological changes in medical images, surpassing CNN-based models in certain domains. However, transformer-based models are required to be trained on very large datasets to achieve optimal performance, which can be a major bottleneck given the scarcity of such datasets in the medical field. Additionally, the high computational costs needed for the multi-head attention module pose practical challenges for real-time applications and deployment in environments with limited computational resources. Furthermore, due to the large number of parameters in transformer-based models, there is an increased risk of overfitting when training on small datasets with subsequent challenges of poor generalization to out-of-domain datasets under such conditions.

\begin{figure*}[!t]
\centering
\includegraphics[width=0.8\textwidth]{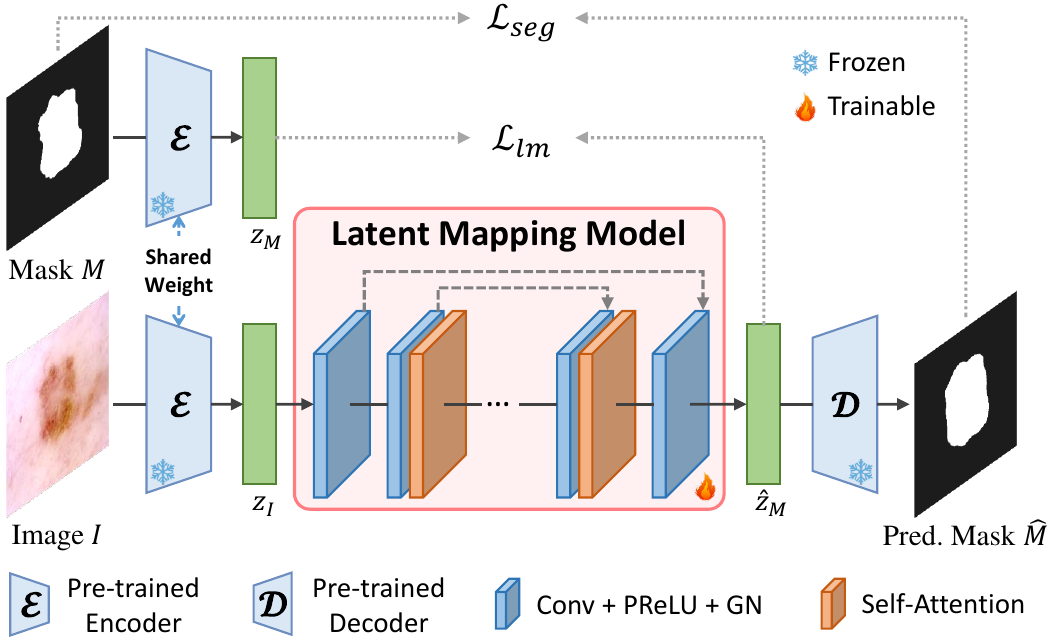} 
\caption{GMS network architecture for 2D medical image segmentation. $\mathcal{E}$ and $\mathcal{D}$ represent a pre-trained vision foundation model and weights are frozen. We utilize the model weights from the Stable Diffusion VAE for $\mathcal{E}$ and $\mathcal{D}$. The latent mapping model (orange box) contains convolution blocks and self-attention blocks but does not contain down-sampling layers. Such a design helps to preserve the spatial information in the input feature vectors. Here, Conv means the 2D convolution operation, and GN represents the Group Normalization.} 
\label{fig:main_framework}
\end{figure*}

Generative models, such as Generative Adversarial Networks (GANs)~\cite{goodfellow2014generative} and Variational Autoencoders (VAEs)~\cite{kingma2013auto}, are often adopted as data augmentation techniques to improve the performance of segmentation models~\cite{huo2022brain}. However, GANs suffer from mode collapse and may distort outputs when the number of training samples is small~\cite{karras2020training}. Additionally, GANs can not guarantee that the distribution of synthetic images is similar to the distribution of real images in the domain. Image-to-image translation models have been used to perform image segmentation in a generative manner, where the image serves as the input and the mask as the prediction. To date, the performance of image-to-image models is well below state-of-the-art segmentation model performance~\cite{li2021semantic}. Recently, MedSegDiff-V2~\cite{wu2024medsegdiff} utilized a diffusion model for medical image segmentation, where a condition model is proposed to encode images into the feature space for mask generation. However, diffusion-based approaches require repetitive denoising steps which lead to longer inference times. GSS~\cite{chen2023generative} is a generative semantic segmentation framework designed for semantic image segmentation, where Vector Quantized Variational Autoencoder (VQ-VAE)~\cite{van2017neural} was employed to project the image and mask into a latent space, and an additional image encoder was designed and trained to match the latent distributions between the mask and image. However, GSS has high computational costs as the additional image encoder is complex, requiring a large number of trainable parameters, to translate the input image into a latent prior distribution.

In this paper, we propose Generative Medical Segmentation (GMS) to perform image segmentation in a generative manner. GMS leverages a pre-trained image encoder to obtain latent representations containing the semantic information for input images and masks, and a latent mapping model is designed to learn a transformation function from the image latent representation to the mask latent representation. The final segmentation mask in the image space is obtained by decoding the transformed mask latent representation using a pre-trained image decoder paired with the pre-trained image encoder. In this approach, only the latent mapping model parameters are learned from the training dataset. The pre-trained image encoder and decoder are obtained from a vision foundation model trained on a large, general dataset. Therefore, the latent representations are more general to unseen data compared to models trained only on images for the desired specific task. We demonstrate GMS achieves the best performance on five public medical image segmentation datasets across different domains. Furthermore, we perform an external validation experiment to demonstrate that the inherent domain generalization ability of GMS is better than other domain generalization methods.

\section{Related Works}
\subsection{Medical Image Segmentation}
Medical image segmentation has experienced rapid advancements in the last decade due to the development of deep-learning techniques. The encoder-decoder architecture with skip connections enables accurate image segmentation by combining low-level and high-level features to perform pixel-wise prediction, making U-Net~\cite{ronneberger2015u} a benchmark method across various medical image segmentation tasks. Subsequent model enhancements such as MultiResUNet~\cite{ibtehaz2020multiresunet} and ACC-UNet~\cite{ibtehaz2023acc} have been implemented using the basic architecture by integrating the residual block or redesigning the hierarchical feature fusion pipeline to gain improved segmentation performance. nnU-Net~\cite{isensee2021nnu} established a guideline for tailoring the receptive field size of convolution kernels and network depth to specific tasks and also incorporated extensive data augmentation into model training to improve segmentation performance for specific medical imaging datasets. 

The Vision Transformer (ViT) introduced the multi-head attention mechanism, which enables capturing long-range feature dependencies across patches in the image, leading to stronger feature representations for image segmentation compared to CNN-based models. This ability to model relationships between distant pixels has proven highly beneficial for medical image segmentation, where understanding the broader imaging context is often crucial to performing the task. ViT-based segmentation models~\cite{cao2022swin,wang2022smeswin,wang2022uctransnet} have competitive results compared against traditional CNN-based models.

\subsection{Generative \& Foundation Models}
Generative models are commonly designed for image synthesis and image-to-image translation tasks. For image synthesis, GANs~\cite{goodfellow2014generative} and VAEs~\cite{kingma2013auto} are often leveraged to generate more data for downstream model training~\cite{huo2022brain,chaitanya2021semi}, especially in the context of medical image segmentation, as the cost of obtaining large, annotated medical imaging datasets is high. Recently, studies have explored the use of diffusion models to create more training instances and alleviate the data scarcity problem~\cite{ye2023synthetic}. However, the iterative denoising process in diffusion models results in a longer inference time compared to GAN or VAE-based approaches. For image-to-image translation, models developed on CNNs~\cite{kong2021breaking} or ViT~\cite{liu2023one} show satisfied results on the MRI missing modality completion task. Currently, few models are designed for performing the image segmentation task directly in a generative manner. GSS~\cite{chen2023generative} is the exception, this model employs VQ-VAE~\cite{van2017neural}  to discretize the image and mask into a finite set of latent codes, which are then reconstructed back into the image space. An independent image encoder is trained to match the image latent codes to the mask latent codes. 

Foundation models, such as Stable Diffusion~\cite{rombach2022high} and Segment Anything (SAM)~\cite{kirillov2023segment}, are trained on large-scale datasets and are designed to generalize across a wide range of tasks. These models are able to serve as a versatile starting point for numerous tasks. Stable Diffusion utilizes a VAE model to first encode the image into the latent space and leverages a UNet to iteratively denoise and reconstruct the latent embeddings, guiding the generation process towards a high-quality output. SAM is designed for image segmentation with a prompt design that allows user interactions to adapt the model to various segmentation tasks with no or minimal fine-tuning. Together, these models exemplify the power and flexibility of foundation models in addressing diverse and complex tasks such as image segmentation.

\section{Methodology}
\subsection{Architecture Overview}
\label{section_framework_overview}
The Generative Medical Segmentation (GMS) model architecture is shown in Figure~\ref{fig:main_framework}. Given a 2D image $I$ and a corresponding segmentation mask $M$, the pre-trained encoder $\mathcal{E}$ is used to obtain latent representations $Z_{I}$ and $Z_{M}$ of $I$ and $M$, respectively. The latent mapping model (LMM) is trained to use $Z_{I}$ to predict an estimated latent representation $\hat{Z}_{M}$ of $M$. $\hat{Z}_{M}$ is then decoded by the pre-trained decoder $\mathcal{D}$ to obtain the predicted segmentation result $\hat{M}$ in the original image space. Note the weights of $\mathcal{E}$ and $\mathcal{D}$ are pre-trained and frozen during both model training and inference, which enables updating only the LMM parameters during training. This approach reduces the number of trainable parameters in the model to be much smaller compared to other state-of-the-art deep learning segmentation models.

\subsection{Image Tokenizer}
The pre-trained encoder $\mathcal{E}$ and decoder $\mathcal{D}$ are treated as an image tokenizer as they map an input from the image space to the latent space ($\mathcal{E}$) or the latent to image space ($\mathcal{D}$). The choice of an appropriate and paired $\mathcal{E}$ and $\mathcal{D}$ to obtain a representative latent space for both input images and masks is critical for GMS performance. In this work, we use the weights of Stable Diffusion (SD) VAE~\cite{rombach2022high} for $\mathcal{E}$ and $\mathcal{D}$. Since SD-VAE was trained on a large natural image dataset~\cite{schuhmann2022laion}, it has a rich and diverse latent information representation, leading to a strong zero-shot generalization ability even for medical images. SD-VAE can achieve near-perfect image reconstruction, which enables the feasibility of training GMS~\cite{rombach2022high}.

SD-VAE is comprised of three down-sampling blocks in $\mathcal{E}$ and three up-sampling blocks in $\mathcal{D}$. The latent representation $Z$ is a 3D tensor containing spatial information ($Z \in \mathbb{R}^{4 \times \frac{H}{8} \times \frac{W}{8}}$ if $I \in \mathbb{R}^{3 \times H \times W}$). Such design enables $Z$ to have a rich feature representation, improved reconstruction quality and enhanced the generalization of the latent representation.

\begin{table*}[!t]
\caption{Quantitative segmentation performance on two ultrasound datasets. The best and second-best performances are bold and underlined, respectively. $^\dagger$ indicates fewer trainable parameters than GMS.}
\label{tab:in_domain_seg_results_bus_busi}
\centering
\fontsize{10}{12}\selectfont
\begin{tabular}{l|l|l|ccc|ccc}
\hline
\multirow{2}*{Type} &\multirow{2}*{Model} &Trainable &\multicolumn{3}{c|}{BUS} &\multicolumn{3}{c}{BUSI} \\
\cline{4-9}
 & &Params (M) &DSC$\uparrow$ &IoU$\uparrow$ &HD95$\downarrow$ &DSC$\uparrow$ &IoU$\uparrow$ &HD95$\downarrow$ \\
\hline
\multirow{5}*{CNN} &UNet               &14.0 &81.50 &70.77 &17.68 &72.27 &63.00 &35.42  \\
                   &MultiResUNet       &7.3  &80.41 &70.33 &19.22 &72.43 &62.59 &34.19  \\
                   &ACC-UNet           &16.8 &83.40 &73.51 &16.49 &77.19 &68.51 &25.49  \\
                   &nnUNet             &20.6 &\underline{85.71} &\underline{78.68} &\underline{11.43} &79.45 &70.99 &\underline{22.13}  \\
                   &EGE-UNet$^\dagger$ &0.05 &72.79 &61.96 &27.73 &75.17 &60.23 &29.51  \\
\hline
\multirow{3}*{Transformer} &SwinUNet       &27.2  &80.37 &69.75 &20.49  &76.06 &66.10 &28.69 \\
                           &SME-SwinUNet   &169.8 &78.87 &67.13 &22.19  &73.93 &62.70 &30.45 \\
                           &UCTransNet     &66.4  &83.44 &73.74 &16.33  &76.55 &67.50 &25.46 \\
\hline
\multirow{4}*{Generative} &MedSegDiff-V2 &129.4 &83.23 &74.36 &17.02  &71.32 &62.73 &38.47  \\
                          &SDSeg         &329.0 &82.47 &73.45 &20.53  &72.76 &63.52 &36.79  \\
                          &GSS           &49.8  &84.86 &77.58 &22.42  &\underline{79.56} &\underline{71.22} &28.20  \\
\cline{2-9}
 &\textbf{GMS (Ours)}                &1.5  &\textbf{88.42} &\textbf{80.56} &\textbf{6.79}  &\textbf{81.43} &\textbf{72.58} &\textbf{19.50}  \\
\hline
\end{tabular}
\end{table*}

\begin{table*}[!t]
\caption{Quantitative segmentation performance on three medical datasets of different modalities. The best and second-best performances are bold and underlined, respectively. $^\dagger$ indicates fewer trainable parameters than GMS.}
\label{tab:in_domain_seg_results_glas_ham_kvasir}
\centering
\fontsize{10}{12}\selectfont
\begin{tabular}{l|l|ccc|ccc|ccc}
\hline
\multirow{2}*{Type} &\multirow{2}*{Model} &\multicolumn{3}{c|}{GlaS} &\multicolumn{3}{c|}{HAM10000} &\multicolumn{3}{c}{Kvasir-Instrument}\\
\cline{3-11}
 & &DSC$\uparrow$ &IoU$\uparrow$ &HD95$\downarrow$ &DSC$\uparrow$ &IoU$\uparrow$ &HD95$\downarrow$ &DSC$\uparrow$ &IoU$\uparrow$ &HD95$\downarrow$ \\
\hline
\multirow{5}*{CNN} &UNet               &87.99 &80.01 &18.45 &92.24 &86.93 &13.74 &93.82 &89.23 &8.71  \\
                   &MultiResUNet       &88.34 &80.34 &17.42 &92.74 &87.60 &13.02 &92.31 &87.03 &9.49  \\
                   &ACC-UNet           &\underline{88.60} &\underline{80.84} &\underline{17.14} &93.20 &88.44 &10.83 &93.91 &89.73 &8.74  \\
                   &nnUNet             &87.25 &78.24 &20.07 &93.83 &\underline{89.32} &\underline{9.43}  &\underline{93.95} &\textbf{90.20} &8.51  \\
                   &EGE-UNet$^\dagger$ &83.25 &71.31 &28.79 &\underline{93.90} &88.50 &10.01 &92.65 &86.30 &9.04  \\
\hline
\multirow{3}*{Transformer} &SwinUNet       &86.44 &76.89 &19.63 &93.51 &88.68 &10.46 &92.02 &85.83 &9.15  \\
                           &SME-SwinUNet   &83.72 &72.77 &26.23 &92.71 &87.21 &12.53 &93.32 &88.27 &8.91  \\
                           &UCTransNet     &87.17 &78.80 &20.79 &93.45 &88.73 &10.91 &93.27 &88.48 &8.84  \\
\hline
\multirow{4}*{Generative} &MedSegDiff-V2 &86.82 &77.05 &19.96 &92.28 &87.02 &13.02 &92.29 &87.21 &9.06 \\
                          &SDSeg         &86.76 &76.23 &21.41 &92.54 &87.53 &12.29 &91.23 &86.54 &9.38 \\
                          &GSS           &87.41 &79.17 &19.81 &92.92 &87.98 &11.29 &93.66 &89.15 &\underline{7.25} \\
\cline{2-11}
 &\textbf{GMS (Ours)}                &\textbf{88.98} &\textbf{81.16} &\textbf{16.32} &\textbf{94.11} &\textbf{89.68} &\textbf{9.32} &\textbf{94.24} &\underline{90.02} &\textbf{7.03} \\
\hline
\end{tabular}
\end{table*}

\subsection{Latent Mapping Model (LMM)}
\label{section_latent_mapping_model}
The latent mapping model (LMM) is the key component in GMS to map from $Z_{I}$ to $Z_{M}$. Instead of using the transformer block that recruits tons of parameters for multi-head attention, we build the LMM with 2D convolutions which make it lightweight. Besides, we do not include any down-sampling layers in LMM to avoid spatial information loss. Note excluding down-sampling layers is not practical in the original UNet model because the receptive fields of the convolution operations are greatly limited if no down-sampling layers are used in the model. Skip connections between convolutional layers are added to prevent vanishing gradients and the loss of semantic-relevant features. 

The model structure is shown in the lower middle of Figure~\ref{fig:main_framework} (orange box). Given $Z_{I}$, which is acquired from the pre-trained encoder $\mathcal{E}$, it first goes through two convolution blocks where each block consists of a 2D convolutional layer (Conv), a PReLU activation function, and group normalization (GN) layer to obtain the feature vector $\mathbf{F}$.

Next, a self-attention mechanism layer is added to better capture global semantic relationships and facilitate feature interaction within $\mathbf{F}$. Specifically, we use three independent convolution operations to generate query $Q$, key $K$, and value $V$, respectively: 
\begin{equation}
Q=W_Q \cdot \mathbf{F} + b_Q, K=W_K \cdot \mathbf{F} + b_K, V=W_V \cdot \mathbf{F} + b_V.
\end{equation}
Here $W$ is the convolution kernel matrix and $b$ is the learnable bias.

Then the self-attention for the query, key, and value is computed as:
\begin{equation}
\mathrm{Attention}(Q, K, V) = \mathrm{softmax}(\frac{QK^T}{\sqrt{d_k}})V, 
\end{equation}
where $d_k$ denotes the feature channel of $K$, and $\mathrm{softmax}$ denotes the softmax normalization function. Due to the small spatial size of the $\mathbf{F}$, employing the self-attention mechanism allows for the efficient capture of long-range dependencies and interactions within the latent representations.

\subsection{Loss Functions}
\label{section_loss_function}
Two loss functions are used to guide model training, a matching loss $\mathcal{L}_{lm}$ in the latent space and a segmentation loss $\mathcal{L}_{seg}$ in the image space. $\mathcal{L}_{lm}$ is formulated to enforce similarity between $Z_{M}$ and $\hat{Z}_{M}$. Specifically, $\mathcal{L}_{lm}$ is defined as:
\begin{equation}
\mathcal{L}_{lm}=\left\|Z_{M}-\hat{Z}_{M}\right\|^2_2.
\label{eq:loss_lm}
\end{equation}
$\mathcal{L}_{seg}$ enforces similarity between the predicted mask $\hat{M}$ and the ground truth mask $M$, even where the latent representation $\hat{Z}_{M}$ deviates from $Z_{M}$. $\mathcal{L}_{seg}$ is defined as: 
\begin{equation}
\mathcal{L}_{seg}=1 - \frac{{2*\sum{M \odot \hat M}}}{{\sum{M+\sum{\hat M}}}},
\label{eq:loss_seg}
\end{equation}
where $\odot$ denotes element-wise multiplication. The final compound loss function used for model training is:
\begin{equation}
\mathcal{L}=\mathcal{L}_{lm}+\mathcal{L}_{seg}.
\label{eq:loss_total}
\end{equation}

\begin{figure*}[!t]
\centering
\includegraphics[width=0.99\textwidth]{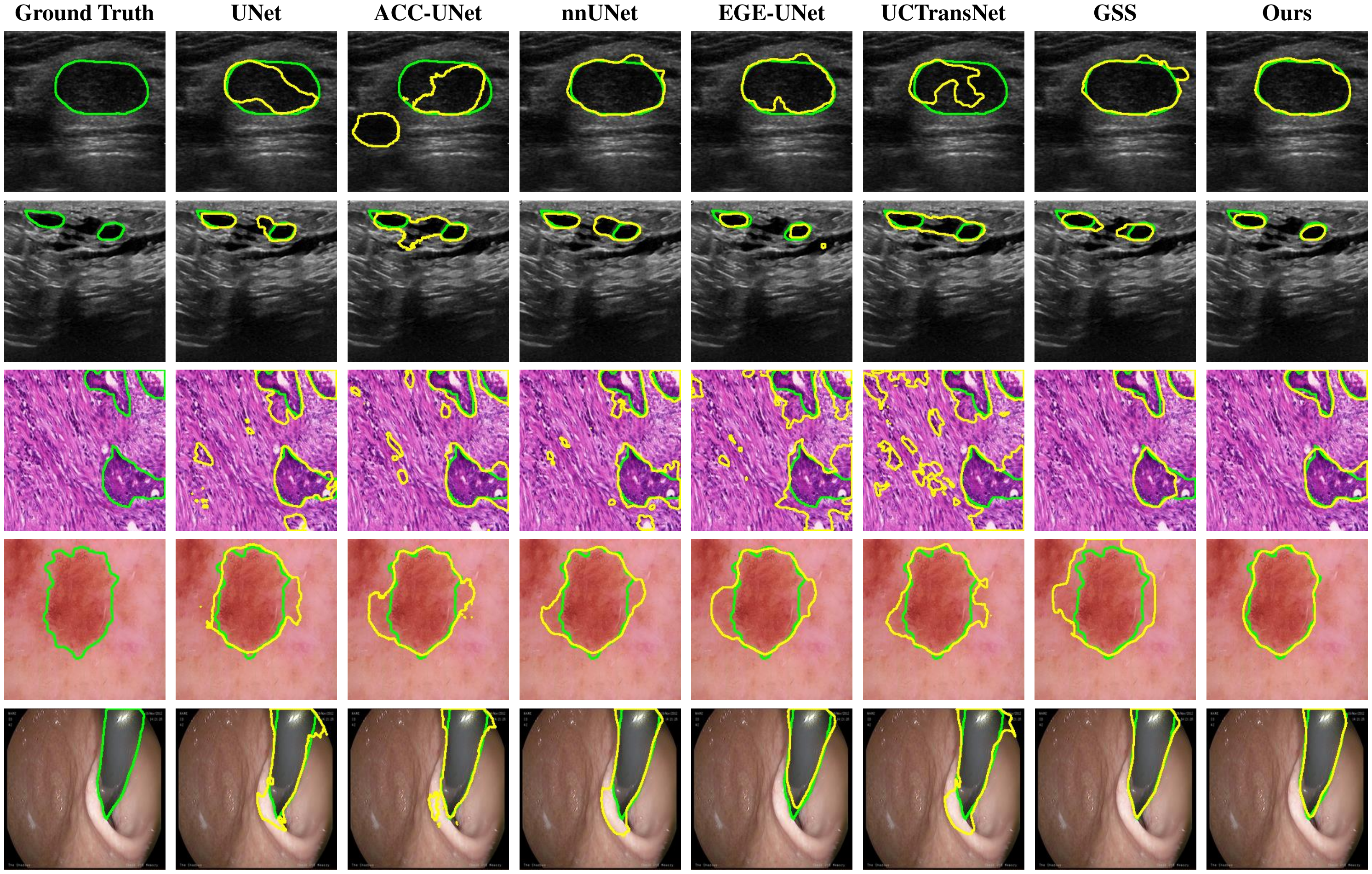} 
\caption{Exemplar segmentation results of GMS and other state-of-the-art methods. From top to bottom are images from the BUS, BUSI, GlaS, HAM10000 and Kvasir-Instrument datasets. The green contours are the ground truth, and the yellow contours are the model predictions. Zoom in for more details.}
\label{fig:seg_visualization}
\end{figure*}

\section{Experiments}
\subsection{Datasets}
We evaluated the performance of GMS on five public datasets: BUS~\cite{yap2017automated}, BUSI~\cite{al2020dataset}, GlaS~\cite{sirinukunwattana2017gland}, HAM10000~\cite{tschandl2018ham10000} and Kvasir-Instrument~\cite{jha2021kvasir}. BUS and BUSI are breast lesion ultrasound datasets that contain 163 and 647 images, respectively. GlaS is a colon histology segmentation challenge dataset divided into 85 images for training and 80 images for testing. HAM10000 is a large dermatoscopic dataset that consists of 10015 images with skin lesion segmentation masks. The Kvasir-Instrument dataset contains 590 endoscopic images with tool segmentation masks. For datasets not already divided, we randomly select 80\% of the images for training and the remaining 20\% for testing.

\subsection{Implementation Details}
Our framework is implemented using PyTorch v1.13, and all model training was performed on an NVIDIA A100 40G GPU. We use AdamW~\cite{loshchilov2018decoupled} as the training optimizer. We utilize the cosine annealing learning rate scheduler to adjust the learning rate in each epoch with the initial learning rate set to $2e^{-3}$. For all experiments, the batch size was set to 8 and the total training epochs were 1000. The input image size is resized to $224\times224$, and on-the-fly data augmentations were performed during training including random flip, random rotation, and color jittering in the HSV domain. We set a threshold of 0.5 to change the predicted gray-scale masks to binary masks. We quantify segmentation performance using the Dice coefficient (DSC), Intersection over Union (IoU), and Hausdorff Distance 95th percentile (HD95).

\subsection{Comparison with State-of-the-Art Models}
We compare GMS with other state-of-the-art methods to evaluate its performance, including CNN-based methods: UNet~\cite{ronneberger2015u}, MultiResUNet~\cite{ibtehaz2020multiresunet}, ACC-UNet~\cite{ibtehaz2023acc}, nnUNet~\cite{isensee2021nnu} and EGE-UNet~\cite{ruan2023ege}; Transformer-based methods: SwinUNet~\cite{cao2022swin}, SME-SwinUNet~\cite{wang2022smeswin} and UCTransNet~\cite{wang2022uctransnet}; and generative methods: MedSegDiff-V2~\cite{wu2024medsegdiff}, SDSeg~\cite{lin2024stable} and GSS~\cite{chen2023generative}). We also compared against two domain generalization models: MixStyle~\cite{zhou2023mixstyle} and DSU~\cite{li2022uncertainty} to evaluate the inherent domain generalization ability of GMS.

\begin{table}[!t]
\caption{Quantitative performance for domain generalization segmentation. A to B indicates A for training and B for testing. Best and second-best performances are bold and underlined, respectively. $^\dagger$ indicates fewer trainable parameters than GMS.}
\label{tab:cross_domain_seg_results}
\centering
\fontsize{10}{12}\selectfont
\begin{tabular}{l|cc|cc}
\hline
\multirow{2}*{Model} &\multicolumn{2}{c|}{BUSI to BUS} &\multicolumn{2}{c}{BUS to BUSI} \\
\cline{2-5}
 &DSC$\uparrow$ &HD95$\downarrow$ &DSC$\uparrow$ &HD95$\downarrow$  \\
\hline
UNet               &62.99 &47.26 &53.83 &96.81  \\
MultiResUNet       &61.53 &53.97 &56.25 &94.31  \\
ACC-UNet           &64.60 &42.87 &47.80 &135.24  \\
nnUNet             &\underline{78.39} &\underline{20.53} &\underline{59.13} &\underline{89.32} \\
EGE-UNet$^\dagger$ &69.04 &34.63 &54.46 &105.23  \\
\hline
SwinUNet       &78.38 &21.94 &57.47 &91.63  \\
SME-SwinUNet   &74.78 &25.81 &58.28 &91.26  \\
UCTransNet     &72.76 &28.47 &56.94 &94.32  \\
\hline
MixStyle       &73.07 &26.52 &57.97 &93.54  \\
DSU            &66.15 &40.03 &56.70 &95.31  \\
\hline
MedSegDiff-V2  &69.56 &32.51 &55.21 &98.57  \\
SDSeg          &74.03 &26.32 &57.03 &94.61  \\
GSS            &68.74 &35.74 &58.72 &92.57  \\
\hline
\textbf{GMS (Ours)}     &\textbf{80.31} &\textbf{18.55} &\textbf{61.60} &\textbf{85.25} \\
\hline
\end{tabular}
\end{table}

Quantitative comparisons of all models on the two ultrasound datasets are presented in Table~\ref{tab:in_domain_seg_results_bus_busi}. GMS achieves the highest DSC, IoU, and HD95 compared to other models. We also present the trainable parameter of each model in Table~\ref{tab:in_domain_seg_results_bus_busi}, where only EGE-UNet has fewer trainable parameters than GMS, and most models have between $\times10$ and $\times100$ more parameters. GMS achieves a 2.71\% and 1.87\% improvement in the DSC metric on the BUS and BUSI datasets, respectively, compared to the second-best model. Additionally, for IoU and HD95 metrics, our approach shows improvement of up to 1.88\% and 4.64, respectively, over the second-best model. Transformer-based segmentation models do not show competitive results on these datasets, which indicates that the intrinsic long-range modeling capability of the transformer block may not be suitable for such tasks as gray-scale ultrasound images lack chromatic information that often aids in distinguishing different tissues. The limited texture and low contrast in these images might reduce the effectiveness of the multi-head attention module in transformer-based models. nnUNet, as a powerful auto-configuration segmentation model, beats transformer-based models and even some generative models, and is the second-best model on the BUS dataset. Notably, segmentation performance is not correlated to the models' number of trainable parameters but is beneficial from the plausible model design and the robust representations provided by the pre-trained vision foundation model. However, models (e.g. EGE-UNet) that contain too few trainable parameters, may lack the capacity to capture complex patterns and relationships in the images, leading to underfitting and poor performance on both datasets. On the contrary, models (e.g. SME-SwinUNet and MedSegDiff-V2) with an excess of parameters can easily overfit the training dataset, memorizing rather than generalizing, which compromises its performance on the test set. 


Table~\ref{tab:in_domain_seg_results_glas_ham_kvasir} presents quantitative results on the other three datasets, where all images are colored. GMS achieves the best segmentation performance except for the IoU metric on the Kvasir-Instrument dataset. It is worth noting that not all generative segmentation models outperform other discriminative models, which proves the importance of design when applying the generative model framework. MedSegDiff-V2 employs an encoder-decoder model to embed images as conditions for guiding the denoising step, yet its performance remains suboptimal. SDSeg utilized the Stable Diffusion model to generate latent representations and further decode them as predicted masks. Additionally, SDSeg proposed a trainable encoder to embed the image into the latent space as the condition for the denoising step. This design does not maximize the use of the knowledge encapsulated in the pre-trained vision foundation model, which may account for its poorer performance. GMS outperforms both CNN-based and transformer-based models, suggesting that generative models when carefully designed can be suitable for a wide variety of segmentation tasks.

\begin{table*}[!t]
\caption{Quantitative segmentation performance on three datasets for ablation study using different loss functions.}
\label{tab:ablation_study_loss}
\centering
\fontsize{10}{12}\selectfont
\begin{tabular}{cc|ccc|ccc|ccc}
\hline
\multirow{2}*{$\mathcal{L}_{lm}$} &\multirow{2}*{$\mathcal{L}_{seg}$} &\multicolumn{3}{c|}{BUSI} &\multicolumn{3}{c|}{HAM10000} &\multicolumn{3}{c}{Kvasir} \\
\cline{3-11}
 & &DSC$\uparrow$ &IoU$\uparrow$ &HD95$\downarrow$ &DSC$\uparrow$ &IoU$\uparrow$ &HD95$\downarrow$ &DSC$\uparrow$ &IoU$\uparrow$ &HD95$\downarrow$ \\
\hline
$\checkmark$ &              &80.25 &71.26 &21.57 &93.92 &89.41 & 9.95 &92.93 &88.28 & 10.21 \\
 & $\checkmark$             &78.75 &69.87 &24.78 &93.64 &88.99 & 10.27 &93.00 &88.47 & 10.68 \\
$\checkmark$ & $\checkmark$ &\textbf{81.43} &\textbf{72.58} &\textbf{19.50} &\textbf{94.11} &\textbf{89.68} &\textbf{9.32} &\textbf{94.24} &\textbf{90.02} &\textbf{7.03}  \\
\hline
\end{tabular}
\end{table*}

\begin{table*}[!t]
\caption{Quantitative segmentation performance on three datasets using different image tokenizers.}
\label{tab:ablation_study_tokenizer}
\centering
\fontsize{10}{12}\selectfont
\begin{tabular}{c|ccc|ccc|ccc}
\hline
\multirow{2}*{Image Tokenizer} &\multicolumn{3}{c|}{BUSI} &\multicolumn{3}{c|}{HAM10000} &\multicolumn{3}{c}{Kvasir} \\
\cline{2-10}
 &DSC$\uparrow$ &IoU$\uparrow$ &HD95$\downarrow$ &DSC$\uparrow$ &IoU$\uparrow$ &HD95$\downarrow$ &DSC$\uparrow$ &IoU$\uparrow$ &HD95$\downarrow$ \\
\hline
VQ-VAE &79.23 &70.34 &24.57  &92.77 &87.61 &13.34  &92.47 &88.31 &9.73  \\
SD-VAE &\textbf{81.43} &\textbf{72.58} &\textbf{19.50} &\textbf{94.11} &\textbf{89.68} &\textbf{9.32} &\textbf{94.24} &\textbf{90.02} &\textbf{7.03}  \\
\hline
\end{tabular}
\end{table*}

\subsection{Domain Generalization Ability}
We evaluated all models on their ability to segment images within the same modality but collected as a part of different datasets to demonstrate model domain generalization ability. Specifically, we train the model using the training set from one dataset but evaluate the performance on a different dataset. This experiment was performed with the BUS and BUSI datasets interchangeably as training and test sets since they are the same modalities (breast ultrasound) but acquired from different centers and vendors. Therefore, the data distributions of the training and test sets are not aligned. Quantitative results are shown in Table~\ref{tab:cross_domain_seg_results}, where GMS outperforms all other models in terms of DSC and HD95. In particular, nnUNet demonstrates powerful domain generalization abilities on both datasets, thanks to its network architecture and the use of data augmentation techniques. However, GMS surpasses nnUNet by around 2\% for DSC. We also recruited two domain generalization methods (MixStyle and DSU) for comparison. MixStyle and DSU were implemented based on DeepLab-V3~\cite{chen2017rethinking} and employed the ResNet50~\cite{he2016deep} as the encoder. GMS is better than two domain generalization methods, which demonstrates the powerful domain generalization ability of our model. The improvements achieved by our model are likely due to the latent representations derived from the pre-trained large vision model, which are domain-agnostic as it was trained on a large, general-purpose dataset. Additionally, GMS has fewer trainable parameters compared to the other generative models, which further reduces the likelihood of overfitting the model to the training set.

\subsection{Qualitative Segmentation Results}
Qualitative segmentation results for different models are shown in Figure~\ref{fig:seg_visualization}. The yellow and green lines denote the contours of predictions and ground truth, respectively. In images acquired from the BUS and BUSI datasets (top two rows), breast lesions show regular shapes, but the segmentation results of other models are often irregular causing over- or under-segmentation. GMS shows the most consistent results compared with the ground truth, which also proves the superiority of GMS. For the histology and dermatology images (the third and fourth rows), there are some regions with highly similar appearances to the target area, which leads to false positive segmentation results in the CNN and transformer-based models. However, GMS is still able to accurately segment those images with complex or misleading patterns. For the endoscopic image (last row), generative methods (GSS and our approach) give the most accurate predictions, which demonstrates the advantages of employing large pre-trained vision models for the segmentation task.

\subsection{Sensitivity Analysis}
\subsubsection{Ablation Studies on Loss Function.}
We performed an ablation study on different loss function combinations for BUSI, HAM10000, and Kvasir-Instrument datasets. As shown in Table~\ref{tab:ablation_study_loss}, the compound loss ($\mathcal{L}_{lm}$ + $\mathcal{L}_{seg}$) always has the best segmentation performance regardless of dataset size or modality. Interestingly, different datasets have different supervision preferences. GMS using only $\mathcal{L}_{lm}$ for model training performs better on BUSI and HAM10000 datasets, which implies supervision in the latent space is more effective compared to the image space. However, GMS performance is better for $\mathcal{L}_{seg}$ when training on the Kvasir-Instrument dataset, indicating supervision in the image space is more important. The compound loss having the best performance suggests that supervision in the image and latent space are both important for achieving the best performance.

\subsubsection{Image Tokenizer Effectiveness.}
We evaluated two pre-trained image tokenizers to assess their performances across three datasets. VQ-VAE~\cite{van2017neural} is a variant of VAE, incorporating a vector quantization step to generate discrete latent representations. Table~\ref{tab:ablation_study_tokenizer} displays the results using VQ-VAE and SD-VAE (default tokenizer used in GMS) as the image tokenizers. SD-VAE improves DSC by up to 2.2\% and reduces the HD95 by up to 5.07, indicating that SD-VAE is more suitable for image tokenization compared to VQ-VAE. The performance improvements also affirm the appropriateness of SD-VAE for handling diverse image segmentation tasks.

\section{Conclusion}
We presented Generative Medical Segmentation (GMS) to perform medical image segmentation. Unlike other segmentation methods where a discriminative model is trained, GMS leverages a powerful pre-trained vision foundation model, Stable Diffusion Variational Autoencoder (SD-VAE), to obtain latent representations of both images and masks. Next, our novel lightweight latent mapping model learns a mapping function from image latent representations to mask latent representations, enabling predicted latent representations of the mask to be generated from an image. Finally, the pre-trained model generates pixel-wise segmentation predictions in the image space using the predicted latent representation of the mask. Experiments on five datasets show that GMS outperforms the state-of-the-art discriminative segmentation models such as ACC-UNet. Moreover, the domain generalization ability of GMS is stronger than other domain generalization models, like DSU and MixStyle, due to the domain-agnostic latent embedding space used by GMS. One key limitation is that GMS can only segment 2D medical images, due to the currently used pre-trained image tokenizer SD-VAE being only trained on 2D natural images. In the future, we will explore extending GMS to 3D medical images by selecting an appropriate pre-trained model for 3D images and adapting the latent mapping model to work on its latent representations.

\bibliography{aaai25}

\end{document}